# Subwavelength Photorefractive Grating in a Thin-Film Lithium Niobate Microcavity

Jiankun Hou[1ǂ], Jiefu Zhu[2ǂ], Ruixin Ma[1], Boyi Xue[1], Yicheng Zhu[1], Jintian Lin[3], Xiaoshun Jiang[4], Xianfeng Chen[2], Ya Cheng[3], Li Ge[5*], Yuanlin Zheng[1,2*] and Wenjie Wan[1,2*]

[1]State Key Laboratory of Advanced Optical Communication Systems and Networks, University of Michigan-Shanghai Jiao Tong University Joint Institute, Shanghai Jiao Tong University, Shanghai 200240, China
[2]School of Physics and Astronomy, Shanghai Jiao Tong University, Shanghai 200240, China
[3]State Key Laboratory of High Field Laser Physics and CAS Center for Excellence in Ultra-Intense Laser Science, Shanghai Institute of Optics and Fine Mechanics, Chinese Academy of Sciences, Shanghai 201800, China
[4]National Laboratory of Solid-State Microstructures, College of Engineering and Applied Science and School of Physics, Nanjing University, Nanjing 210093, China
[5]Department of Physics and Astronomy, College of Staten Island, the City University of New York, NY 10314, USA
ǂ *Equal contribution*
*Corresponding authors*: Wenjie Wan wenjie.wan@sjtu.edu.cn or or Yuanlin Zheng ylzheng@sjtu.edu.cn

## Abstract

Subwavelength gratings play a fundamental and pivotal role in numerous science and applications for wave manipulation, exhibiting distinctive features such as filtering, phase manipulation, and anti-reflection. However, conventional fabrication methods for ultrasmall periodic structures are constrained by the fundamental optical diffraction limit, making it challenging to produce subwavelength gratings for optics. Here, we demonstrate a novel technique to build a reconfigurable subwavelength photorefractive grating (SPG) in a thin-film lithium niobate on the platform of an optical microcavity. Such SPGs are optically induced through the photorefractive effect and the subwavelength features originate from the spatial phase modulations of the pump's standing wave. The resulting SPGs lead to the mode splitting of two counter-propagating modes inside the microcavity, exhibiting an Electromagnetically Induced Transparency (EIT)-like transmission spectrum. Moreover, the unique subwavelength characteristic of SPGs enables first-order quasi-phase-matching for backward second-harmonic generation, a long-standing problem in nonlinear optics. Also, free-space-to-chip vertical nonlinear frequency conversion can be achieved in a similar manner. These results provide a flexible approach towards fabricating subwavelength gratings, which holds significant potential in various applications such as nonlinear frequency conversion, optical communication, sensing, and quantum technologies.

## Introduction

Optical gratings, a fundamental optical element with a periodic structure, exhibit distinct responses to the incoming light field's wavelengths, enabling a wide range of applications in spectroscopy, dielectric mirror coating, pulse shaping, wavelength multiplexing, and nonlinear periodically poled crystals[1-3]. These coherent and collected dynamics from multiple periods in one dimension (1D) later are extended into 2D, and even 3D to develop the prominent photonic band theory[4,5]. Critically, an optical grating with subwavelength periodicity can further manipulate the flow of light at the sub-micro scale, enabling exotic behaviors like ultra-narrow Bragg reflection[6], negative refraction[7], light localization, and even super-resolution imaging[8]. One intuitive way to understand the subwavelength's nature is within the momentum space, where the linear optical processes should fulfill the rigid conservation law, while the subwavelength gratings can provide the required large wave vectors for various purposes. However, to realize such subwavelength gratings, the fabrications demand high-precision optical coating[9] or high-resolution photolithography[10], making it an extremely challenging task.

In nonlinear optics, one form of grating structures, namely the periodic poling technique, is widely used to compensate the momentum mismatching to achieve quasi-phase-matching (QPM) for improving the efficiency of nonlinear frequency conversion in the medium[11,12]. Several approaches have been demonstrated for QPM, such as photogalvanic-field-induced all-optical poling in silicon nitride photonics[13-16], hyperpolarizability of intermolecular charge transfer states in organic semiconductors[17], or periodically poled domain inversion in ferroelectric crystals[18]. In recent years, periodically poled lithium niobate (PPLN) has been extensively used in planar-integrated photonic devices for on-chip nonlinear processes[19-21]. At the chip scale, it is meaningful to explore the subwavelength grating for QPM, as it is capable of unlocking some schemes that the traditional QPM with longer periods can impossibly provide. The implied nonlinear processes like mirrorless optical parametric oscillation (OPO)[22], second-harmonic generation (SHG) in the ultra-violet wavelength range[23], and backward second-harmonic generation (BSHG)[24], counter-propagating spontaneous parametric down-conversion (SPDC)[25] are critical in their practical applications. However, achieving subwavelength periods remains a challenge for conventional periodic poling methods due to the stringent requirement on poling periods over a subwavelength distance. Although there have been reports on the successful fabrication of sub-micro periodic domains (or stacked metasurfaces)[17,22,26-28], the fabrication of subwavelength gratings for QPM in ferroelectric crystals has still been challenging.

Here, we experimentally demonstrate a novel technique for constructing a reconfigurable SPG within a thin-film lithium niobate (TFLN) optical microcavity. In contrast to the conventional periodic domain inversion, our SPG is fabricated using two counterpropagating pumps via the photorefractive effect. The spatial phase of the grating automatically aligns with that of the standing wave through the electro-optic effect due to the internal space charge field. This unique SPG significantly enhances the coupling between the clockwise (CW) and counterclockwise (CCW) modes, leading to a reconfigurable mode splitting and an EIT-like transmission spectrum. Furthermore, we utilize the short-period QPM grating to enable BSHG through inter-mode coupling. The output BSHG exhibits a linear correlation with the strength of the grating. Finally, we demonstrate the vertical excited SHG from free space to the microcavity. In contrast to the conventional double resonant SHG process between confined modes in a microcavity, this free-space-to-chip nonlinear coupling only depends on the grating's wave vector and requires second harmonic modes in resonance solely. Our work introduces a novel physical mechanism for achieving subwavelength gratings and holds significance for on-chip photonic nonlinear conversion devices for all-optical signal processing and quantum information.

## Results

**SPG formation in LN microcavity**

Lithium niobate (LN) presents a very intriguing combination of electro-optic (EO) properties and nonlinear-optic characteristics, making it highly promising for fabricating high-performance integrated photonic devices. As a ferroelectric material, LN undergoes refractive index changes based on the spatial distribution of light intensity when illuminated, known as the photorefractive effect[29]. In recent decades, this effect has effectively been harnessed for applications such as holographic storage[30], phase-conjugate mirror[31], and information processing in bulk crystals[32]. Furthermore, this effect has also been observed in high-quality (Q) LN microresonators with a significant boost from enhanced intracavity optical intensity. Another notable characteristic of LN is the large second-order nonlinear susceptibility $\chi^{(2)}$, which serves as the foundation for nonlinear optical processes such as SHG, sum/difference frequency generation (SFG/DFG), OPO, and SPDC. Combining both features of LN's photorefractivity and second-order nonlinearity could hold the key for efficient nonlinear conversion.

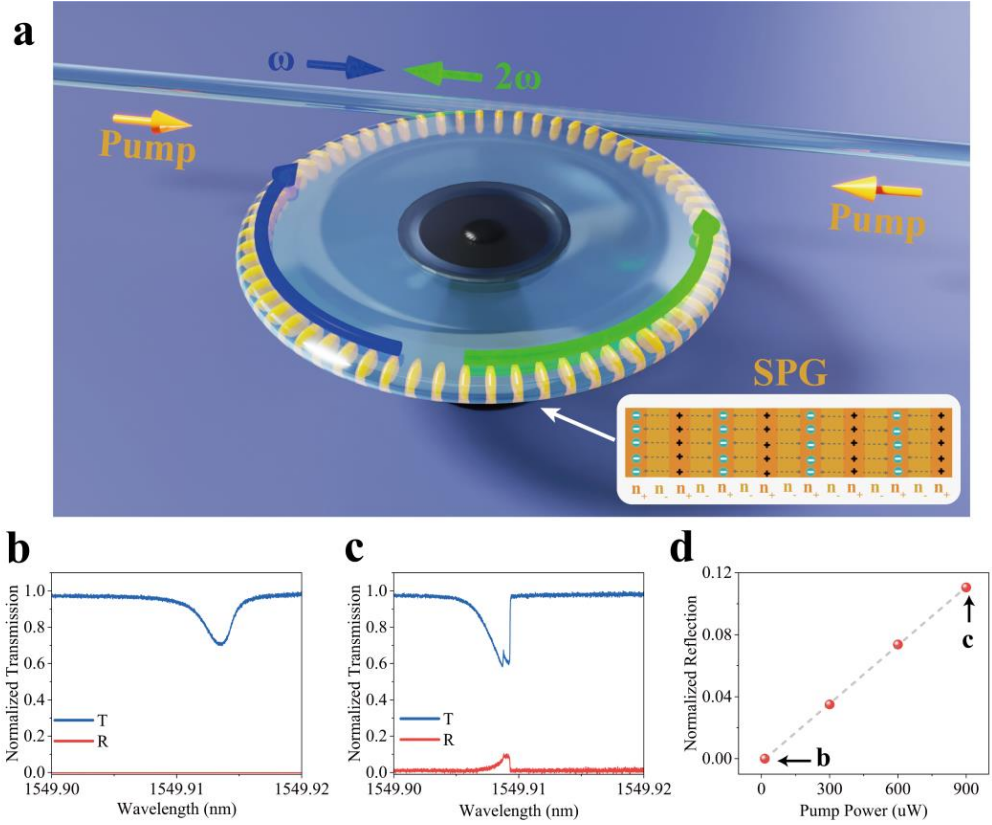

**Fig. 1. Illustration of SPG within a TFLN microcavity. a** Schematic of a tapered fiber coupled TFLN microcavity with an SPG inscribed by standing optical wave. The SPG is induced by counterpropagating pumps (yellow) and provides a long vector to satisfy QPM for BSHG. Inset: charge distribution. The transmission spectra of FW show almost no reflection without grating in **b** and about 12% back-reflection at high pump power in **c**, where an SPG has been established in the microcavity. The measurements are normalized according to the transmission. **d** The grating strength has a linear dependence on the pump power, expressed in terms of the reflection of the input FW light.

Experimental realization of SPG consists of a z-cut undoped high-Q ($\sim 10^6$) TFLN microdisk resonator with a diameter of 100 μm and a thickness of 600 nm, fabricated from commercial TFLN using UV photolithography and chemo-mechanical polishing (CMP)[33,34]. As depicted in Fig. 1a, counter-propagating pumps are evanescently coupled to the microresonator via a tapered fiber to excite whispering gallery modes. Evidently, these counter-propagating waves form a standing wave pattern characterized by zero amplitude at nodes and maximum amplitude at antinodes. As a photorefractive material, the intense irradiation on specific regions of the microresonator leads to ionization and subsequent charge distribution, resulting in the generation of a space charge field in LN. Meanwhile, the internal space charge field induces changes in the refractive index of the microresonator within areas where the field is most pronounced through an electro-optic effect[35]. Consequently, a refractive index grating with a spatial period approximately half that of the pump light wavelength emerges (Fig. 1a inset). Moreover, due to the high

Q factor and cavity resonance enhancement, the presence of SPG can be evidently observed in our TFLN microdisk despite the relatively small refractive index modulation ($\Delta n \sim 10^{-9}$)[36]. This SPG possesses a large grating vector, enabling some intriguing nonlinear effects such as BSHG and vertical excited SHG, which shall be elaborated on later.

Theoretically, the dynamics of the system can be described by the temporal coupled-mode theory (TCMT), as given:

$$\frac{d}{dt}a_1 = \left(i\Delta - \frac{\gamma}{2}\right)a_1 + i\kappa_g a_2 + i\sqrt{\gamma_E}s_{in}, \qquad (1)$$

$$\frac{d}{dt}a_2 = \left(i\Delta - \frac{\gamma}{2}\right)a_2 + i\kappa_g a_1, \qquad (2)$$

where $a_1$ and $a_2$ are amplitudes of the CW and CCW modes, respectively. $\Delta = \omega - \omega_0$ denotes the detuning from the resonant mode, $s_{in}$ is the incident light field, and $\gamma_E$ represents the coupling strength with which the cavity is coupled to the waveguide. The effective loss in the resonator is defined as $\gamma = \gamma_E + \gamma_0$, where $\gamma_0$ is the intrinsic loss term which is dominated by material absorption and surface impurities. The coupling term $\kappa_g$ between the CW and CCW modes arises from both resonator inhomogeneities (material or geometry) and the induced photorefractive grating, which could be expressed as [36]:

$$\kappa_g(t) = \kappa_0 + \xi\left(1 - e^{-t/\tau_p}\right)I_p, \qquad (3)$$

where $\kappa_0$ is the scattering from static inhomogeneities. The induced grating emerges endogenously from the field ionization of space charges that diffuse over a timescale $\tau_p$ to generate a local charge density. The resulting static electric field induces the SPG via the electro-optic effect. After a long time ($t \gg \tau_p$), the effective depth of the SPG is controlled by the coefficient $\xi$ and the power of the input pump $I_p$. Moreover, this SPG has a subwavelength period $\Lambda = \lambda_p/2n_{eff}$, where $\lambda_p$ is the pump wavelength and $n_{eff}$ is its effective refractive index, which corresponds to the intensity distribution of the standing wave.

## Generation of subwavelength photorefractive grating

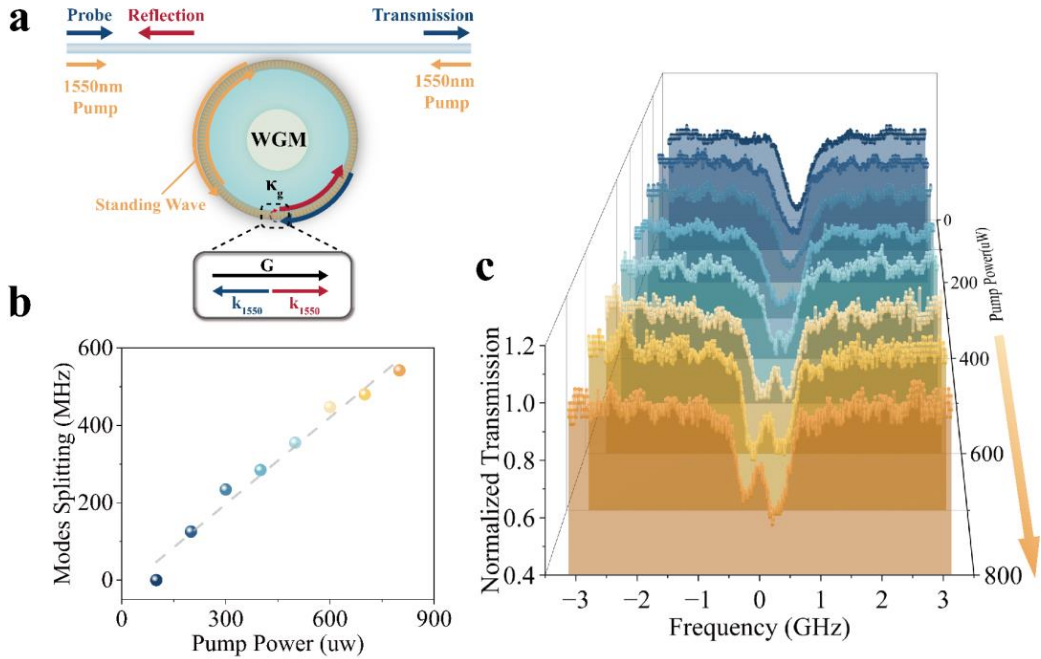

**Fig. 2. EIT-like transmission of the SPG in the microresonator. a** Schematic diagram of the SPG induced by a 1550 nm pump entering from both sides simultaneously. CW and CCW modes are coupled through SPG with a coupling strength $\kappa_g$. Mode splitting is measured using a 1550 nm probe with a lock-in amplifier. **b** and **c** Experimental data of the mode splitting process for different pump powers. The dashed line shows the linear dependence of the splitting on the pump power. **c** EIT-like transmission spectrum resulted from the CW and CCW mode splitting.

Firstly, we investigate the behavior of a stably produced SPG ($t \gg \tau_p$). Since the SPG has a period equal to half the pump wavelength, it inherently acts as a Bragg reflection grating for the pump. The CW and CCW modes at the reflected wavelength ($\lambda_{\text{pump}}$) are coupled to each other assisted by the SPG, manifesting it as mode splitting under strong coupling. As shown in Fig. 1b, at a counterpropagating pump power of 15 μW, only minimal ionization-induced free electron generation takes place, thus photorefractive effect-induced refractive index change is negligible and no reflection or normal transmission spectra are observed. However, when the pump power increases to 900 μW, the photorefractive effect is greatly enhanced and produces a grating with sufficient strength to induce mode coupling between the CW and CCW modes. Fig. 1c captures the immediate aftermath of deactivating one of the input ports, where approximately 12% of the light is reflected into the backward propagating mode by the induced grating, accompanied by a clearly observable frequency splitting. The observed blue shift is attributed to a decrease in the refractive index resulting from the photorefractive effect (Supplementary

Material). In addition, we find that the strength of the SPG is linearly dependent on the pump power and could be quantified in terms of reflection as depicted in Fig. 1d.

The mode splitting is also observed to increase with the pump power. To measure the precise mode splitting without the thermal-pulling effect, we employ a lock-in amplification technique on the probe beam[37], which detects the presence of any dynamical mode indicated by Eqns. (1) and (2). Meanwhile, we manage to input a 1550-nm pump from both ports and precisely tune the pump's power using variable optical attenuators (VOA), allowing us to control the strength of the SPG, as depicted in Fig. 2a. Further details are given in supplementary. As previously mentioned, it is worth noting that the SPG can be considered as a Bragg grating for inducing pumps, thereby reflecting CW light to its counterpart, CCW light. When the SPG is in a stable state, we slowly scan the probe at a speed of $0.01\ nm/s$ near the pump resonance mode from the blue-tuned region. Subsequently, we observe mode splitting in the transmission spectra due to SPG-induced modal coupling between the CW and CCW traveling modes (Fig. 2c). As the pump laser power gradually increases, starting from a few microwatts, a distinct mode splitting becomes evident in the through-port, indicating that the strength of the SPG is directly proportional to the pump's power. At low pump power levels, we consider the broadening of the resonant mode as indicative of mode splitting where it is difficult to discern. However, at high pump power levels, we can determine that the mode splitting corresponds to the scattering rate $2\kappa_g$[38], reaching 550 MHz when using an 800 µW pump laser (as shown in Fig. 2b) and resulting in an EIT-like transmission spectrum[39]. The Q factors of the split modes remain constant throughout the process, implying that the mode-coupling mechanism has a negligible impact on resonator losses. It is worth noting that the photorefractive effect does not exhibit wavelength selectivity, thus theoretically enabling SPG generation for all resonant modes in microcavities.

**Backward second harmonic generation induced by SPG**

QPM is a widely employed technique for compensating the momentum mismatch caused by dispersion, commonly utilized in forward second-harmonic generation (FSHG) where both the pump and SHG propagate in the same direction. The momentum mismatch given by $\Lambda_{\text{FSHG}} = \frac{2\pi}{\Delta k} = \frac{\lambda}{2|n_{2\omega}-n_\omega|}$ can be conveniently satisfied using PPLN, where $\Delta k$ represents the phase mismatch, $\lambda$ is the fundamental wavelength (FW), $n_{2\omega}$ and $n_\omega$ correspond to the effective refractive indices at SH and FW, respectively.

However, attaining BSHG necessitates a sub-micron QPM period ($\Lambda_{BSHG} = \frac{2\pi}{\Delta k} = \frac{\lambda}{2|n_{2\omega}+n_{\omega}|}$), which still poses challenges in terms of manufacturing.

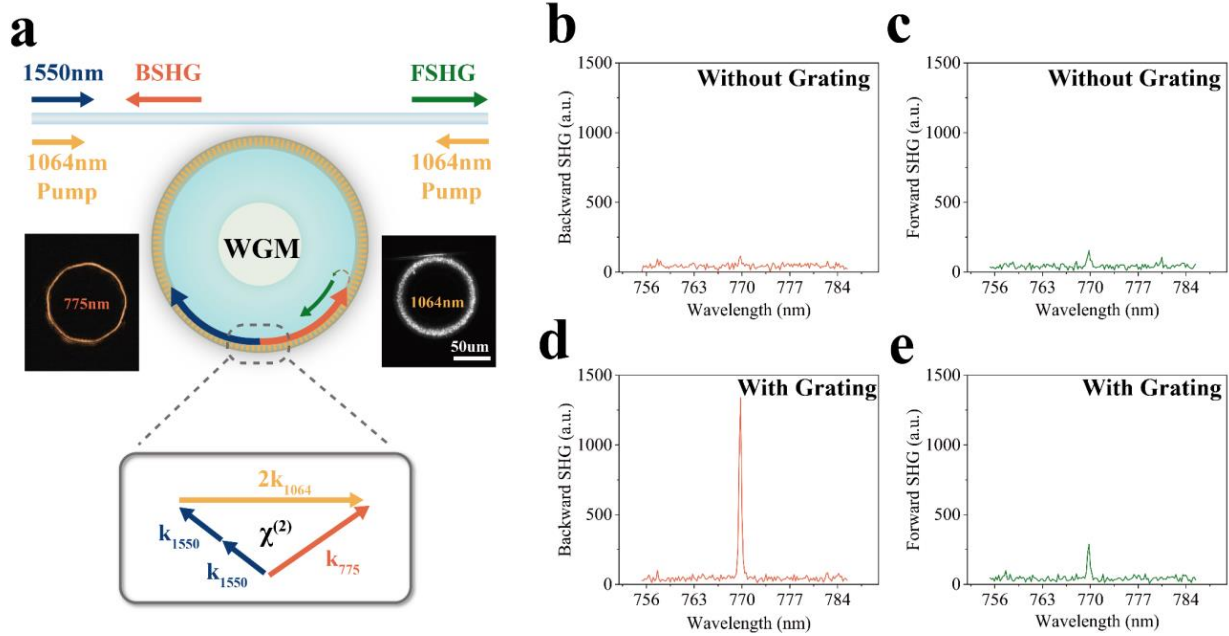

**Fig. 3. BSHG induced by SPG in a TFLN microresonator. a** Schematic diagram of the BSHG process in the microcavity. The bottom shows a phase-matching scheme of BSHG involving a lower-order 1064nm mode and a higher-order 775nm mode. Left inset: high-order polygon mode near 775 nm. Right inset: fundamental mode for 1064 nm. **b** and **c** Both BSHG and FSHG are negligible without pump input. However, after construction of the SPG, a strong BSHG can be observed and some forward SHG can be detected in **d** and **e**.

As a proof of principle, we demonstrate subwavelength QPM gratings induced by the photorefractive effect to enable BSHG. A tunable continuous wave laser at 1064 nm is employed as the pump source and its output is amplified before being injected from both ports of the tapered waveguide for evanescent coupling of light into the TFLN microcavity. This injection results in the formation of a standing wave around the resonator circumference with a spatial period of $\Lambda = \lambda_{1064}/2n_{1064} \approx 266$ nm, as illustrated in Fig. 3a. Simultaneously, FW at 1550 nm is introduced through the left port to excite the CW mode in the microcavity. To satisfy the QPM condition $\Delta k = 2k_{1064} = k_{775} - 2k_{1550}$ for BSHG, we select the fundamental mode for the pump and higher-order mode for FW and SH. The left inset in Fig. 3a shows the higher-order polygon mode for BSHG and the right inset displays the fundamental mode for 1064 nm (The image for FW mode was unable to capture only due to our experimental constraints). The resonant mode for FW is analyzed from its transmission spectrum. Considering the short period of SPG, only the $m = -1$ diffraction order is allowed for SHG according to the nonlinear grating equation[17]. Then, BSHG and FSHG signals are separated by WDMs connected to each port and finally detected by spectrometers (details provided in Supplementary Materials). Figs. 3b and 3c depict the spectra of BSHG and FSHG in the absence of SPG, i.e., no pump input, respectively. The phase mismatched SHG only generates negligible signals in CW and CCW directions. Conversely, upon introducing a pump power of 7 mW and achieving stable build-up of the SPG, a robust generation of BSHG occurs, corresponding to an enhancement of $10^2$, with no significant alteration observed in FSHG (only 2 times enhancement). The slight increase in FSHG may be attributed to the backscattering from BSHG. It should be noted that to avoid affecting the SPG throughout the experiment, the power of the FW is maintained at 200 µW.

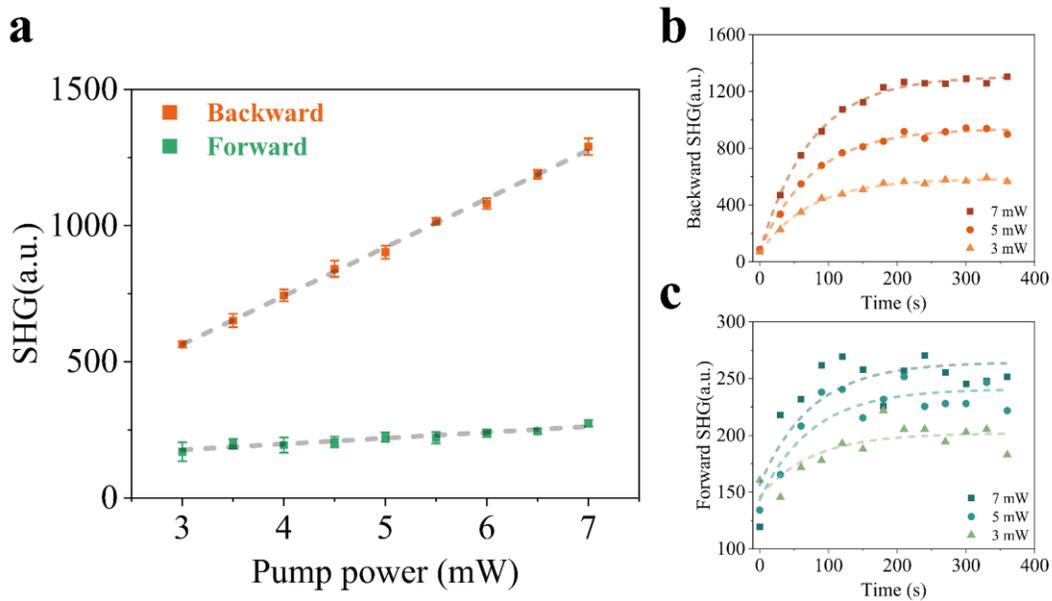

**Fig. 4. Dynamic response of SHG in the TFLN microresonator via photorefraction induced SPG. a** The SHG intensity shows a linear relationship with the grating intensity, i.e., the pump power. The dashed grey lines are the linear fitting. The error bars represent the standard deviation of five repeated measurements. **b** and **c** Experimental measurements of BSHG and FSHG over time at different pump powers, respectively. Both curves can be fitted to exponential responses (dashed curves) with $\tau \approx 75s$.

To investigate the correlation between SHG and SPG, we examine the intensity of BSHG and FSHG under varying input pump powers, i.e., different strengths of SPG. In our experimental setup, we adjust the VOA to modulate the pump input power while maintaining a fixed FW input power at 200 µW. Remarkably, both BSHG and FSHG exhibit an exponential increase in intensity towards a steady-state value, characterized by a substantial time constant associated with the photorefractive effect[40,41]. Figs. 4b and 4c show the temporal behavior of BSHG and FSHG, respectively, at three distinct pump powers. These results demonstrate a consistent time constant of ~75 s, irrespective of the input pump power.

Notably, the result also reveals that FSHG primarily arises from the backscattering of BSHG induced by QPM originating from SPG. Moreover, we observe a positive linear correlation between SHG and pump power (SPG strength), as depicted in Fig. 4a, which aligns with theoretical predictions. Thus, dynamic control over BSHG can be achieved by adjusting the pump input power to modulate the strength of SPG.

**Vertical excited SHG**

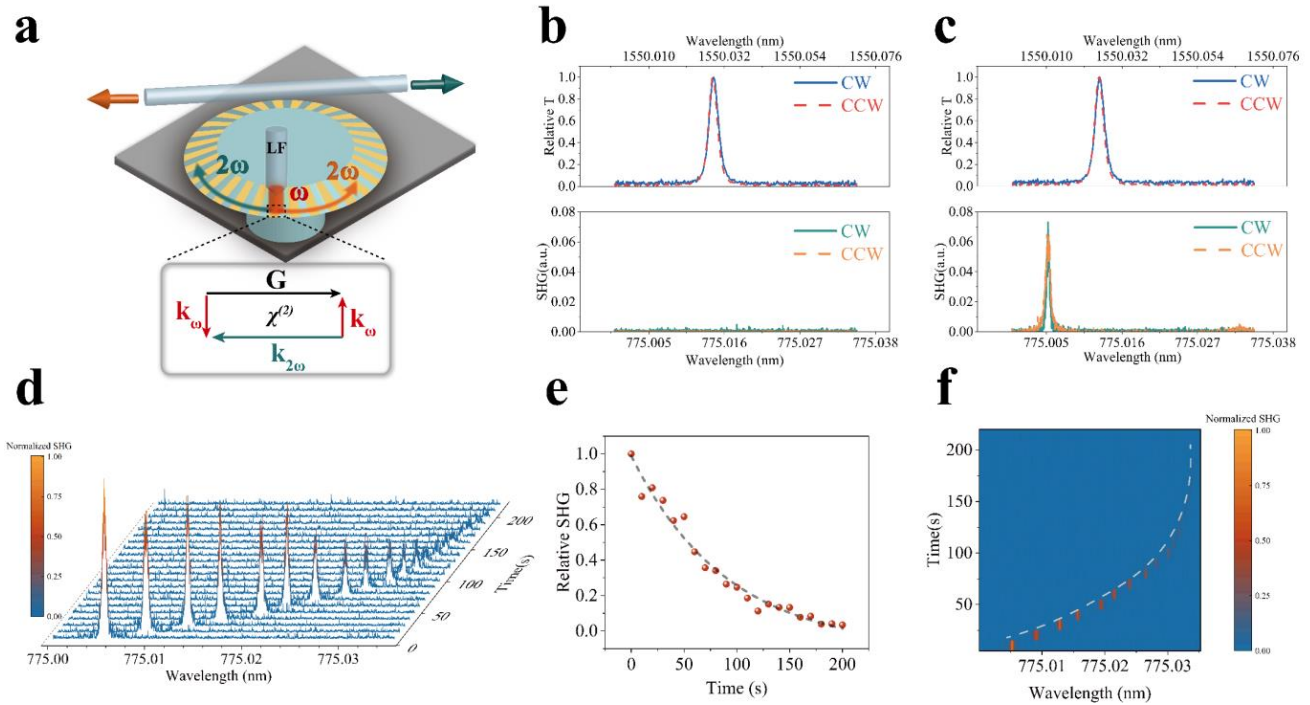

**Fig. 5. Vertical excited SHG process. a** Schematic illustration of the vertical excited SHG process. The FW at frequency $\omega$ is incident vertically on the surface of a TFLN microdisk via a lens fiber. The orange and green arrows indicate the counterpropagating SHG at $2\omega$. SPG is induced by a standing wave from the 1547.1 nm pump. The inset shows the momentum conservation. **b** Only FW is observed from the output ports without SPG. **c** SHG signal is detected from both ports after the establishment of SPG. The separation of the resonance wavelength is due to dispersion. **d-f** The dynamic process of SHG during the photorefractive decay when the pump is switched off. Both the intensity and the detuning will have an exponential response with $\tau \approx 90$ s. The intensity of SHG is normalized to the maximum value. FW: 10 mW. Pump: 20 mW.

The nonlinear optical radiation and its reversal free-space-to-chip nonlinear frequency conversion of integrated photonic devices have been reported recently[42]. In this section, we use the SPG to implement vertical excited SHG from free space to the microdisk. To accomplish this, we meticulously design and execute an experiment involving vertical injection of FW at 1550 nm into the microdisk through a lens fiber, as schematically illustrated in Fig. 5a. The SPG is induced by a counterpropagating pump at 1547.1 nm and provides a grating vector $G$. Then, two vertically input telecom photons are converted to one visible photon propagating in the microdisk with the momentum conservation $|k_{2\omega}| = |G|$. The process is only efficient when the SH is resonant in the microcavity. Dispersion gives rise to the result that the experimentally obtained FW and pump wavelengths are different. FW and SH signals are evanescently coupled out by a tapered fiber, separated by two WDMs connected to each port, and finally detected by PDs and PMTs, respectively. More details about the experimental setup are provided in the Supplementary Materials.

Fig. 5b shows the output spectra without SPG as the wavelengths are finely tuned, and only the FW signal hitting the resonant mode is detected. When the SPG is built up, SHG signals at the resonant mode are also observed, as shown in Fig. 5c. CW and CCW signals are equal due to the smooth surface and the symmetry of the microdisk. In addition, these two resonant modes are separated in wavelength, which means that the SH signals are generated from nonlinear diffraction through SPG, not by the up-conversion of fundamental resonant modes. Finally, we look at the dynamic process of SHG after switching off the pump. In Figs. 5d-5f, we can see that when the pump laser is switched off, the SHG decays with a time constant of about 90s, and at the same time the resonant mode moves to the longer wavelength with the same time constant, corresponding to the decay of the SPG. This observation opens new avenues for nonlinear diffraction by subwavelength gratings.

**Discussion**

In summary, we have experimentally demonstrated a reconfigurable SPG formed in a high-Q TFLN microdisk. This SPG benefits from the photorefractive effect with a short period equal to half of the wavelength of the incident light. Experimentally, we observe a mode splitting of CW/CCW modes induced by the SPG. Such an SPG also provides a large grating vector enabling a backward QPM scheme. We have shown the BSHG with the assistance of an SPG inscribed using counterpropagating pumps, which has potential in nonlinear on-chip integrated photonics. Besides, this SPG can implement free-space-to-chip vertical nonlinear frequency conversion. The SHG efficiency is proven to be directly dependent on the strength of the SPG via the photorefractive effect. In the future, we can design a microring structure to increase mode overlap or use a femtosecond laser to enhance the photorefractive effect to improve the performance[26]. Our results provide a promising approach to constructing reconfigurable subwavelength gratings without complex fabrication processes. The mechanism is universal and can be extended to other photorefractive materials and structures. The concept can be further applied to large-scale integrated applications, such as integrated compact optical vortex beam emitters[43], photonic isolators[44], vertically emitting lasers[45] and all-optical memories[46,47].

## Methods

**Resonator and tapered fiber fabrication**

The TFLN microdisk resonator is fabricated from a commercial 600 nm thick z-cut TFLN wafer (NANOLN) using UV photolithography-assisted chemo-mechanical etching technology (CMP) in four steps[33]. Firstly, a thin layer of Cr with a thickness of 900 nm is deposited on the surface of the TFLN by thermal evaporation deposition. Subsequently, the Cr film is patterned into a circular disk via UV photolithography and wet etching. Next, a CMP process is performed using a polishing machine (NUIPOL802, Kejing, Inc.), velvet cloth, and polishing slurry (MasterMet, 60 nm amorphous colloidal silica suspension). Then, a TFLN microdisk is defined. The residual Cr film is removed by a Cr etching solution. A secondary CMP is performed to further smoothen the surface of the microdisk. Finally, a buffered HF solution to partially remove the underneath $SiO_2$ layer to form the pillar supporting the microdisk. The wedge angle is controlled by the time of CMP. In our experiment, the thickness of the microdisk is 600 nm and the wedge angle is ~10°.

The tapered fiber is fabricated by heating a standard single-mode silica fiber (G.652.D) with a hydrogen flame while pulling it apart axially (the heat-and-pull technique). Within the coupling region, the tapered fiber has a diameter of 2~3 μm and maintains contact with the microdisk resonator throughout the experiment to ensure stability.

**Experimental details**

In this experiment, a continuously tunable laser (CTL, Toptica Photonics, 1510-1630 nm) is used as the fundamental wave or probe. A low-noise single-frequency fiber laser (NKT Y-10) at 1064 nm is used as the pump laser in BSHG. In the vertical excited SHG experiment, a second continuously tunable laser

(New Focus, TLB-6728) is used as the 1550-nm pump. The phase of the CW and CCW pump needs to be the same to maintain a stable standing wave, otherwise, SPG cannot be formed. When the relative phase is not stable, the SPG is "blurred" and the effect is averaged out. This is solved by adding a delay line in one of the branches to make sure the optical paths of both are equal. If the phases of the two branches are the same, there is no dense oscillation in the transmission spectrum[48].

To obtain mode splitting spectra of the microresonator mode, the probe laser is scanned by voltage modulation with a triangle wave signal at 20 Hz, and its power is modulated by an electro-optic modulator (EOM) at 300 kHz. The probe is transformed into an electrical signal by a PD after passing through the system, and then a lock-in amplifier collects signals at the same scanning frequency. the corresponding setup is shown in Supplementary Materials. Polarisation controllers (PC) are used for optimizing the coupling to each resonance mode. Two 1064/1550-nm and 1064/780-nm wavelength division multiplexers (WDM) are used to separate the signals of different wavelengths in the tapered fiber for detection. Two power meters are used to monitor the input power of each branch.

The lens fiber is precisely positioned using a three-axis piezo nano stage (Thorlabs, MDT630B), with ~10 nm resolution along the vertical direction. The output signals reach the maximum when the gap between the microresonator and lens fiber tip falls into working distance, that is when the FW is focused into the microdisk.

The SH signal is detected by a VIS-NIR spectrometer (Figs. 3 and 4) or photomultiplier tube (PMT, CR114) connected to an OSC (Fig. 5). The transmission spectra of FW and pump are detected by PDs (DET02AFC) while sweeping the laser wavelength. A 1064/780-nm WDM is used to filter out the 1064-nm pump signal and ensures that the residual 1064-nm signals do not affect the results.

**Acknowledgment**: This work was supported by the National key research and development program (Grant No. 2023YFB3906400, No. 2023YFA1407200). National Science Foundation of China (Grant No. 12274295, No. 92050113).

**Author contributions:**

W.W. initiated the idea and designed the study; Y. Z. performed the theoretical study; W.W., Y. Z, X.C., Y. C. supervised the work; J. H. performed experimental work; J. Z., J. L. fabricated the devices; R. M., B. X., Y. Z. helped analyze the data and discussion; W.W., Y. Z., J. H. wrote the paper; All authors reviewed the manuscript.

**Conflict of interest:**

The authors declare no conflict of interest.

**Data and materials availability:**

The data that support this article are available from the corresponding author upon reasonable request.